\documentclass[prc,superscriptaddress,unsortedaddress,twocolumn,showpacs]{revtex4}
\usepackage[dvipdfmx]{graphicx}
\usepackage{graphicx}
\usepackage{amsmath}
\usepackage{amssymb}
\usepackage{times}
\usepackage{bm}
\usepackage{mediabb}
\usepackage{color}
\usepackage{ulem}

\def\la{\mathrel{\mathpalette\fun <}}

\def\fun#1#2{\lower3.6pt\vbox{\baselineskip0pt\lineskip.9pt
\ialign{$\mathsurround=0pt#1\hfil##\hfil$\crcr#2\crcr\sim\crcr}}}
\begin{document}

\title{ Investigating the alpha-clustering on the surface of $^{120}$Sn
via ($p$,$p\alpha$) reaction
and the validity of the factorization approximation
}

\author{Kazuki Yoshida}
\email[]{yoshidak@rcnp.osaka-u.ac.jp}
\affiliation{Research Center for Nuclear Physics (RCNP), Osaka
University, Ibaraki 567-0047, Japan}

\author{Kosho Minomo}
\affiliation{Research Center for Nuclear Physics (RCNP), Osaka
University, Ibaraki 567-0047, Japan}

\author{Kazuyuki Ogata}
\affiliation{Research Center for Nuclear Physics (RCNP), Osaka
University, Ibaraki 567-0047, Japan}

\date{\today}

\begin{abstract}
The $^{120}$Sn($p$,$p\alpha$)$^{116}$Cd reaction at 392 MeV
is investigated with the distorted wave impulse approximation (DWIA) framework.
We show that this reaction is very peripheral mainly because of the strong
absorption of $\alpha$ by the reaction
residue $^{116}$Cd, and the $\alpha$-clustering on the
nuclear surface can be probed clearly.
We investigate also the validity of the so-called factorization approximation
that has frequently been used so far.
It is shown that the kinematics of $\alpha$ in the nuclear interior region
is significantly affected by the distortion of $^{116}$Cd, but it has
no effect on the reaction observables because of the strong absorption in
that region.
\end{abstract}

\pacs{24.10.Eq, 25.40.-h, 27.60.+j}
%24.10.Eq Coupled-channel and distorted-wave models
%25.40.-h Nucleon-induced reactions
%27.60.+j 90<A<149

\maketitle

\section{Introduction}
Nuclear clustering has been one of the main subjects
in nuclear physics; for a recent review, see Ref.~\cite{Hor12}.
As a new topic, $\alpha$-clustering on the surface of heavy nuclei, Sn isotopes,
is theoretically predicted in Ref.~\cite{Typ14}.
This result itself is interesting and important because it has been believed that $\alpha$-clustering is developed mainly in light nuclei, although some indication for $\alpha$-clustering in $^{40}$Ca and $^{44}$Ti was discussed~\cite{Yam98}.
Furthermore, the
result gives a significant impact
on the nuclear equation of state~\cite{Typ14}.

As emphasized in Ref.~\cite{Hor12}, however, one should keep it in mind
that a large spectroscopic factor of $\alpha$
does not necessarily indicate the $\alpha$-clustering, because of
the {\it duality} of the mean-field-type structure and the cluster structure~\cite{Bay58}.
On the other hand, the localization of $\alpha$ in the nuclear surface region is direct evidence of the $\alpha$-clustering in a nucleus.
In this view, the $\alpha$ transfer reaction, ($^6$Li,$d$) in particular,
has been utilized for investigating the $\alpha$-clustering.
Very recently, a three-body reaction model with a
microscopic cluster wave function was applied to the $^{16}$O($^6$Li,$d$)$^{20}$Ne
reaction and the transfer cross section was shown to be sensitive to the
$\alpha$ distribution on the nuclear surface of $^{20}$Ne~\cite{Fuk16}.

In the present study, as an alternative way to the $\alpha$ transfer
reaction, we consider the proton-induced $\alpha$ knockout reaction on
$^{120}$Sn, and investigate how clearly
it can probe the $\alpha$
distribution in the surface region of $^{120}$Sn, i.e., the $\alpha$-clustering of $^{120}$Sn.
We adopt the distorted wave impulse approximation (DWIA)
framework to describe the ($p$,$p\alpha$) reaction; DWIA
has successfully been used in the analysis of
various nucleon knockout~\cite{Sam86,Cha77,Sam87,Jac66,Jac73,Kit85}
and $\alpha$ knockout~\cite{Car84,Mab09,Roo77,Nad80,Nad89,Wan85,Yos88}
experiments.
In many preceding studies, however, the so-called factorization
approximation, which factors out the nucleon-nucleon ($NN$)
transition amplitude in the evaluation of the total transition matrix
element of the knockout process, has been
adopted.
In this paper we explicitly examine the validity of the
factorization approximation by means of the local semi classical approximation
(LSCA)~\cite{Luo91,Wat00} to the distorted waves.
It was argued in Ref.~\cite{Sam86} that the factorization approximation
becomes questionable when the distortion effect is large. It is thus
important to examine its validity for the $\alpha$ knockout process for a heavy nucleus,
in which the distortion on $\alpha$ by the reaction residue
is expected to be very strong.

In Sec.~\ref{secform} we describe the DWIA formalism
for the ($p$,$p\alpha$) reaction, with
introducing the LSCA
that is a key prescription for discussing the accuracy of
the factorization approximation.
In Sec.~\ref{secresult} 
first we show the comparison between the present calculation 
and the experimental data. Next
we discuss
the validity of the factorization approximation
in the $^{120}$Sn($p$,$p\alpha$)$^{116}$Cd reaction at 392~MeV.
We then show that the $^{120}$Sn($p$,$p\alpha$)$^{116}$Cd reaction probes the
$\alpha$ distribution in the surface region with high selectivity.
The dependence of these findings on the $\alpha$ wave function is
also discussed.
Finally, a summary is given in Sec.~\ref{secsum}.

\section{Formalism}
\label{secform}
We consider the A($p$,$p\alpha$)B reaction in normal kinematics in the DWIA framework.
The incoming proton in the initial channel
is labelled as particle 0, and
the outgoing proton and $\alpha$ are particles 1 and 2, respectively.
A (B) denotes the
target (residual) nucleus.
${\bm K}_i$ and $\Omega_i$ ($i=0,1,2$) represent the momentum and its solid angle, respectively, and
$E_i$ ($T_i$) is the total (kinetic) energy of particle $i$.
All quantities with and without superscript L indicate that they are
evaluated in the laboratory (L) and center-of-mass (c.m.) frame, respectively.

\begin{figure}[htbp]
\begin{center}
 \includegraphics[width=0.35\textwidth]{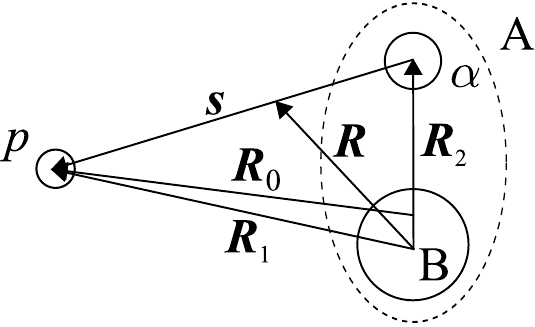}
 \caption{
 Coordinates of A($p$,$p\alpha$)B reaction.
 }
 \label{figcoord}
\end{center}
\end{figure}

The transition amplitude in the DWIA formalism is given by
\begin{align}
&T^{nljm}_{{\bm K}_0{\bm K}_1{\bm K}_2}
= \nonumber \\
&\left<
\chi_{1,{\bm K}_1}^{(-)}({\bm R}_1)\chi_{2,{\bm K}_2}^{(-)} ({\bm R}_2)
\left| t_{p\alpha}({\bm s}) \right|
\chi_{0,{\bm K}_0}^{(+)}({\bm R}_0) \varphi_{\alpha}^{nljm}({\bm R}_2)
 \right>,
 \label{kotmtx}
\end{align}
where $\chi_0$, $\chi_1$, and $\chi_2$ are the scattering
wave functions of the $p$-A, $p$-B, and $\alpha$-B systems, respectively, $t_{p\alpha}$
is the transition interaction between $p$ and $\alpha$,
and $\varphi_{\alpha}^{nljm}$
is the $\alpha$-cluster wave function.
$n$, $l$, $j$, and $m$ are, respectively, the principal quantum number,
the orbital angular momentum, the total angular momentum, and its third component of $\alpha$ in the nucleus A.
The superscripts $(+)$ and $(-)$ specify the outgoing and incoming
boundary conditions on $\chi_i$, respectively.
The definition of the coordinates is given in Fig.~\ref{figcoord}.

By using
\begin{align}
{\bm R}&=\frac{1}{A_\alpha+1}{\bm R}_1
+
\frac{A_\alpha}{A_\alpha+1}{\bm R}_2, \\
{\bm s}&={\bm R}_1-{\bm R}_2,
\end{align}
${\bm R}_i$ are
written by
\begin{align}
{\bm R}_0&={\bm R}_1-\frac{A_\alpha}{A}{\bm R}_2 \nonumber \\
&=\left(1-\frac{A_\alpha}{A}
\right){\bm R}
+\alpha_0\frac{A_\alpha}{A_\alpha+1}{\bm s}, \\
{\bm R}_1&={\bm R}+\frac{A_\alpha}{A_\alpha+1}{\bm s}, \\
{\bm R}_2&={\bm R}-\frac{1}{A_\alpha+1}{\bm s},
\label{r2}
\end{align}
where $A_\alpha=4$ and $\alpha_0=(A+1)/A$
with $A$ the mass number of A.
We make the LSCA~\cite{Luo91,Wat00} that
describes the propagation of the scattering wave for a short distance
$\Delta R$ by a plane wave, i.e.,
\begin{align}
\chi_{i,{\bm K}_i}({\bm R}+\Delta{\bm R})\approx
\chi_{i,{\bm K}_i}({\bm R})\,e^{i{\bm K}_i({\bm R})\cdot \Delta{\bm R}}.
\label{eqlsca}
\end{align}
The norm of the local momentum ${\bm K}_i({\bm R})$ is given by
\begin{align}
|{\bm K}_i({\bm R})|&={\rm Re}[{\bm K}_{i}^{\rm C}({\bm R})],
\end{align}
where
the complex
momentum ${\bm K}_{i}^{\rm C}({\bm R})$ is determined
so as to satisfy the local energy conservation:
\begin{align}
\frac{\left(\hbar  {\bm K}_{i}\right) ^2}{2\mu_i}
=\frac{\left(\hbar {\bm K}_{i}^{\rm C}({\bm R})\right)^2}{2\mu_i}+U_i({\bm R})
\end{align}
with
$\mu_i$ and $U_i({\bm R})$ the reduced mass
of the scattering particles and the distorting
potential for particle $i$, respectively.
The direction of ${\bm K}_i({\bm R})$ is taken to be parallel to the flux of $\chi_{i,{\bm K}_i}({\bm R})$.
The validity of the LSCA is discussed in Sec.~\ref{seclsca}.

Equation~(\ref{kotmtx}) is then reduced to
\begin{align}
T_{{\bm K}_0 {\bm K}_1 {\bm K}_2}^{nljm}
\approx
&\int d{\bm R}\, F_{{\bm K}_0 {\bm K}_1 {\bm K}_2}({\bm R})\,\varphi_{\alpha}^{nljm}({\bm R}) \nonumber \\
&\times
\tilde{t}_{p\alpha}({\bm \kappa'}({\bm R}),{\bm \kappa}({\bm R})),
\label{redtmat}
\end{align}
where $F_{{\bm K}_0 {\bm K}_1 {\bm K}_2}({\bm R})$ and
$\tilde{t}_{p\alpha}({\bm \kappa'}({\bm R}),{\bm \kappa}({\bm R}))$ are defined by
\begin{align}
F_{{\bm K}_0 {\bm K}_1 {\bm K}_2}({\bm R})
&\equiv
\chi_{1,{\bm K}_1}^{*(-)}({\bm R})\, \chi_{2,{\bm K}_2}^{*(-)}({\bm R}) \nonumber \\
&\, \times \chi_{0,{\bm K}_0}^{(+)}({\bm R})\,
e^{-i{\bm K}_0 ({\bm R})\cdot {\bm R}A_\alpha/A}, \\
\tilde{t}_{p\alpha}({\bm \kappa'}({\bm R}),{\bm \kappa}({\bm R}))
&\equiv
\int d{\bm s}\,e^{-i{\bm \kappa'}({\bm R})\cdot{\bm s}}\,
t_{p\alpha}({\bm s})\,e^{i{\bm \kappa}({\bm R})\cdot{\bm s}}.
\end{align}
Here, ${\bm \kappa}({\bm R})$ (${\bm \kappa'}({\bm R})$)
is the $p$-$\alpha$
relative momentum in the initial (final) channel:
\begin{align}
{\bm \kappa}({\bm R})
&\equiv\alpha_0 \frac{A_\alpha}{A_\alpha+1}{\bm K}_0({\bm R})
-\frac{1}{A_\alpha+1}{\bm K}_{\alpha}({\bm R}), \\
{\bm \kappa'}({\bm R})
&\equiv\frac{A_\alpha}{A_\alpha+1}{\bm K}_1({\bm R})
-\frac{1}{A_\alpha +1}{\bm K}_2({\bm R}).
\end{align}
${\bm K}_{\alpha}({\bm R})$ is determined by
the momentum conservation of the $p$-$\alpha$
system:
\begin{align}
{\bm K}_\alpha({\bm R})={\bm K}_1({\bm R})
+{\bm K}_2({\bm R})-\alpha_0{\bm K}_0({\bm R}).
\end{align}

In taking the squared modulus of Eq.~(\ref{redtmat}), we make the
on-the-energy-shell (on-shell) approximation to $\tilde{t}_{p\alpha}$:
\begin{align}
\frac{\mu_{p\alpha} ^2}{(2\pi\hbar ^2)^2}
|\tilde{t}_{p\alpha}({\bm \kappa'}({\bm R}),{\bm \kappa}({\bm R}))|^2
\approx
\frac{d\sigma _{p\alpha}}{d\Omega _{p\alpha}}
(\theta _{p\alpha}({\bm R}),E_{p\alpha}({\bm R})),
\label{onshell}
\end{align}
where $\theta_{p\alpha}({\bm R})$ is the angle between
${\bm \kappa}({\bm R})$ and ${\bm \kappa'}({\bm R})$, i.e., the
local $p$-$\alpha$ scattering angle, and $E_{p\alpha}({\bm R})$ is the
local scattering energy
defined by
\begin{align}
E_{p\alpha}({\bm R})
&=\frac{\hbar ^2 {({\bm \kappa}'({\bm R}) )}^2}{2\mu_{p\alpha}}.
\label{locale}
\end{align}
In Eqs.~(\ref{onshell}) and (\ref{locale}) $\mu_{p\alpha}$ is
the reduced mass of the $p$-$\alpha$ system.

With the LSCA and the on-shell approximation, the triple differential cross section (TDX) of the ($p,p\alpha$) reaction is given by
\begin{align}
\frac{d^3 \sigma}{dE_1^{\rm L} d\Omega_1^{\rm L} d\Omega_2^{\rm L}}
=S_{\alpha}F_{\rm kin}C_0
\sum_m \left| \bar{T}^{nljm}_{{\bm K}_0{\bm K}_1{\bm K}_2} \right| ^2,
\end{align}
where $S_{\alpha}$ is the spectroscopic factor of the alpha-cluster and
the kinematical factor $F_{\rm kin}$ is defined by
\begin{align}
F_{\rm kin}&\equiv J_{\rm L}\frac{K_1 K_2 E_1 E_2}{\hbar ^4 c^4}
\left[1+\frac{E_2}{E_{\rm B}}+\frac{E_2}{E_{\rm B}}
\frac{{\bm K}_1 \cdot {\bm K}_2}{K_2 ^2}\right]^{-1}
\end{align}
with $J_{\rm L}$ the Jacobian from the c.m frame to the L frame, and
\begin{align}
C_0&=\frac{E_0}{(hc)^2 K_0}\frac{1}{(2\ell+1)}\frac{\hbar ^4}{(2\pi)^3 \mu_{p\alpha}^2}.
\end{align}
The reduced transition amplitude is given by
\begin{align}
\bar{T}^{nljm}_{{\bm K}_0{\bm K}_1{\bm K}_2}
=&\int d{\bm R}\,
\sqrt{
\frac{d\sigma _{p\alpha}}{d\Omega _{p\alpha}}
(\theta _{p\alpha}({\bm R}),E_{p\alpha}({\bm R}))
} \nonumber \\
&\times F_{{\bm K}_0 {\bm K}_1 {\bm K}_2}({\bm R})\,\varphi_{\alpha}^{nljm}({\bm R}).
\label{localtmt}
\end{align}

In the preceding studies on knockout reactions~\cite{Sam86,Cha77,Sam87,Jac66,Jac73,Kit85}, further simplification of $\bar{T}^{nljm}_{{\bm K}_0{\bm K}_1{\bm K}_2}$
was made by replacing ${\bm K}_i({\bm R})$ with the asymptotic momentum
${\bm K}_i$. We then obtain
\begin{align}
\frac{d^3 \sigma}{dE_1^{\rm L} d\Omega_1^{\rm L} d\Omega_2^{\rm L}}
\rightarrow
\,&
F_{\rm kin}C_0
\frac{d\sigma _{p\alpha}}{d\Omega _{p\alpha}}(\theta _{p\alpha},E_{p\alpha})
\nonumber \\
&
\times \sum_m \left|
\int d{\bm R}\,F_{{\bm K}_0 {\bm K}_1 {\bm K}_2}({\bm R})\,\varphi_{\alpha}^{njlm}({\bm R})
\right| ^2,
\label{asymptmt}
\end{align}
where $\theta _{p\alpha}$ and $E_{p\alpha}$ are given in the
same way as for $\theta _{p\alpha}({\bm R})$ and $E_{p\alpha}({\bm R})$,
respectively, but with using the asymptotic $p$-$\alpha$ relative momenta:
\begin{align}
{\bm \kappa}
&
\equiv\alpha_0 \frac{A_\alpha}{A_\alpha+1}{\bm K}_0
-\frac{1}{A_\alpha+1}{\bm K}_{\alpha},
\\
{\bm \kappa'}
&
\equiv\frac{A_\alpha}{A_\alpha+1}{\bm K}_1
-\frac{1}{A_\alpha +1}{\bm K}_2.
\end{align}
This prescription is called the factorization approximation.
One sees that this approximation is equivalent to use
the asymptotic momentum ${\bm K}_i$ instead of
the local momentum ${\bm K}_i({\bm R})$ in Eq.~(\ref{eqlsca}),
i.e.,
\begin{align}
\chi_{i,{\bm K}_i}({\bm R}+\Delta{\bm R})\approx
\chi_{i,{\bm K}_i}({\bm R})\,e^{{\bm K}_i\cdot \Delta{\bm R}},
\label{eqama}
\end{align}
which we call the asymptotic momentum approximation (AMA).
Therefore the accuracy of the factorization approximation can be judged,
in principle, by that of the AMA.

\section{Results and discussion}
\label{secresult}
\subsection{Numerical inputs}
\label{subsecinput}
For the bound state wave function $\varphi_{\alpha}^{nljm}$, we assume that
the $\alpha$ particle is bound in the 4$S$ orbit in a Woods-Saxon
potential
$V(R)=V_0/(1+{\rm exp}[(R-r_0A^{1/3})/a_0])$
with $r_0=1.27$~fm  and $a_0=0.67$~fm.
The depth of the potential $V_0$ is adjusted so as to reproduce
the $\alpha$ separation energy of $^{120}$Sn, 4.81 MeV.
In the calculation shown below,
the $\alpha$ spectroscopic factor $S_\alpha$ for $^{120}$Sn is taken to be
0.022~\cite{Jan79}.
It should be noted that the purpose of the present study is not to
determine $S_\alpha$ but to understand the property of the $(p,p\alpha)$
knockout reaction and to examine the reliability of DWIA with the factorization
approximation.

\begin{figure}[htbp]
\begin{center}
 \includegraphics[width=0.45\textwidth]{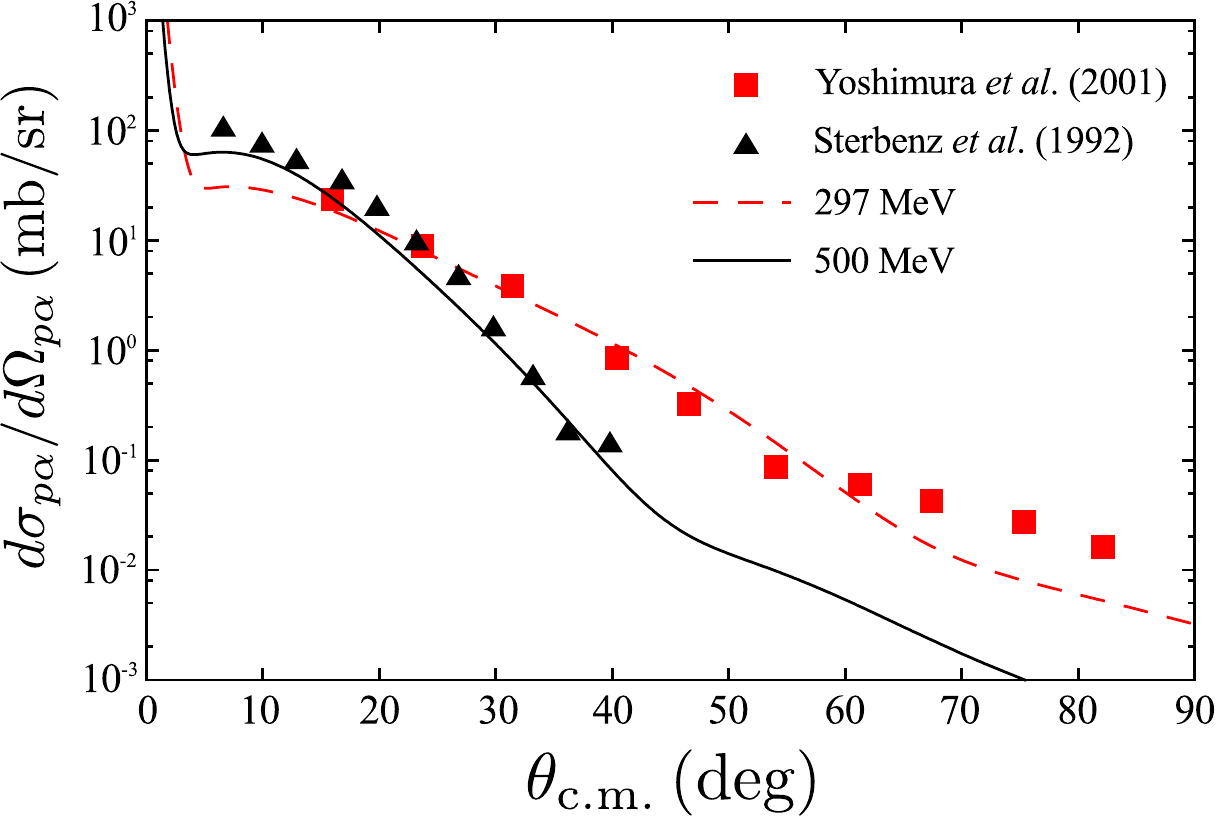}
 \caption{
  (Color online) Comparison between the d$\sigma_{p\alpha}/d\Omega_{p\alpha}$ calculated by the single folding model calculation
  and the experimental data~\cite{Yos01,Ste92} at 297~MeV and 500~MeV.
  }
 \label{figelm}
\end{center}
\end{figure}

One of the most important ingredients of the present DWIA is the
$p$-$\alpha$ differential cross section
$d\sigma_{p\alpha}/ d\Omega_{p\alpha}$ that determines the transition
strength of the $(p,p\alpha)$ process. Because $d\Omega_{p\alpha}$
for various scattering energies and angles are needed, we adopt
the microscopic single folding model~\cite{Toy13} with implementing the
phenomenological nuclear density of $\alpha$ and the
Melbourne $NN$ $g$-matrix interaction~\cite{Amo00}.
As shown in Fig~\ref{figelm}, with no free parameter,
the calculated $d\sigma_{p\alpha}/d\Omega_{p\alpha}$
agrees very well with the experimental data~\cite{Yos01,Ste92}
at 297~MeV and 500~MeV.

As for the distorting potential for $\alpha$ in the final channel,
for consistency, we employ the double folding model~\cite{Ega14}
using the same ingredients as used in the $p$-$\alpha$ calculation;
we use the nuclear density of $^{116}$Cd calculated by the Hartree-Fock
method in the same way as Ref.~\cite{Min10}.
It is known that to phenomenologically determine a low-energy scattering potential
of $\alpha$ is quite difficult because of the discrete ambiguities~\cite{Nol87,Lou78}.
In fact, there have been many attempts
\cite{Ohk04,Cuo10,Fur06} to microscopically determine an $\alpha$ potential
with the double-folding model approach.
It should be noted, however, that in the present
study we evaluate both the real and imaginary parts of the $\alpha$ potential
with no free adjustable parameter, in contrast to those preceding studies.
For the distorting potential of proton in the initial and final channels,
we use the EDAD1 parameter set of the Dirac phenomenology~\cite{Ham90Coo93}.
The Coulomb terms of the distorting potentials are constructed by
assuming that the target (residual) nucleus is a uniformly charged sphere
with the radius of $r_0A^{1/3}$ ($r_0B^{1/3}$).

The effect of the nonlocality of the proton and alpha distorting potentials
is taken into account by multiplying the scattering waves by the Perey
factor~\cite{Per62}
$F_{P}(R)=[1-\mu \beta ^2/(2\hbar ^2)U(R)]^{-1/2}$,
where $\mu$ is the reduced mass between the two scattering particles.
The range of nonlocality $\beta$
for $p$ ($\alpha$) is taken to be 0.85~fm (0.2~fm)~\cite{ManTWOFNR}.

We take the following kinematical condition on the
$^{120}$Sn($p$,$p\alpha$)$^{116}$Cd reaction at 392~MeV;
the Madison convention is adopted.
The kinetic energy of particle 1 is fixed at 328~MeV and
its emission angle is set to $(\theta_1,\phi_1)=(43.2^\circ,0^\circ)$.
As for particle 2, $\phi_2$ is fixed at $180^\circ$ and $\theta_2$
is varied around $61^\circ$; the kinetic energy $T_2$ changes around
59~MeV and
$\theta_{p \alpha} \sim 56^\circ$, $E_{p\alpha}\sim 385$ MeV,
accordingly~\cite{Ues}. We always adopt the relativistic kinematics for all
the scattering particles in this study.

\subsection{Test of the present calculation}
\label{sectest}
\begin{figure}[tbp]
\begin{center}
 \includegraphics[width=0.35\textwidth]{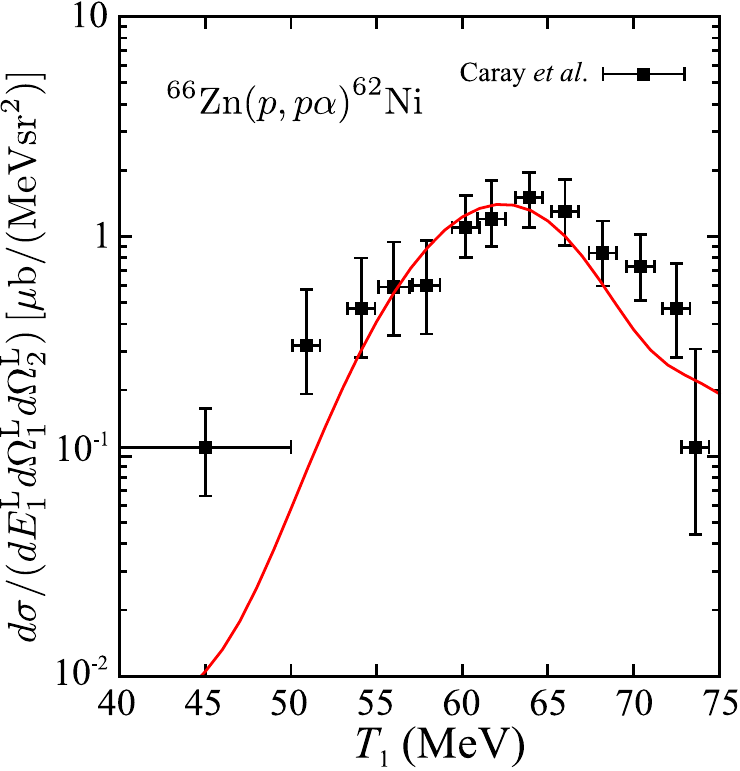}
\end{center}
 \caption{
 (Color online)
  Calculated energy sharing cross section of $^{66}$Zn($p$,$p\alpha$)$^{62}$Ni reaction at 101.5~MeV.
  The experimental data are from Ref.~\cite{Car84}.
 }
 \label{fig3}
\end{figure}

We test the present model calculation by comparing the calculated result of
the energy sharing cross section, which is
a TDX with fixed $d\Omega_1^{\rm L}$ and $d\Omega_2^{\rm L}$, as a function of $T_1$ for
$^{66}$Zn($p$,$p\alpha$)$^{62}$Ni reaction
with measured experimental data~\cite{Car84}; the incident energy is 101.5~MeV.
The present result and the experimental data are shown in Fig.~\ref{fig3}.
The EDAD parameter set are used for the distorting potential of $p$-$^{66}$Zn
and $p$-$^{62}$Ni, and the double folding model is adopted for $\alpha$-$^{62}$Ni,
in the same way as in \ref{subsecinput}.
According to Ref.~\cite{Car84}, we assume that the $\alpha$ particle is bound in the 6$S$ state
in a Woods-Saxon potential with $r_0$=1.30 fm and $a_0=0.67$~fm, and
the depth of the potential $V_0$ is adjusted
so as to reproduce the $\alpha$ separation energy 4.58 MeV.

One can see that the present calculation  well reproduces the observed
energy sharing cross section; the deduced $\alpha$ spectroscopic factor is 0.84, which is sizably larger than the value 0.42 obtained in
the previous study~\cite{Car84}.
It should be noted, however, that the double folding model for the distorting potential of $\alpha$-$^{62}$Ni
will have some ambiguities due to the relatively low scattering energy of $T_2 \sim 30$ MeV.
Furthermore, the calculated result in Ref.~\cite{Car84} showed
quite large ambiguities ($\sim 50$\%) of the deduced $\alpha$ spectroscopic factors due to the $\alpha$-$^{62}$Ni potential.
Considering these facts, it can be concluded that the present result is consistent with the
experimental data and its analysis.

\subsection{Validity of the LSCA and the AMA}
\label{seclsca}

The validity of the LSCA for the scattering of nucleon has been examined
in Refs.~\cite{Wat00,Min10} and it was concluded that at energies
higher than about 50~MeV, the LSCA works for the propagation within
1.5~fm. Furthermore, at those energies the AMA is found to work at almost
the same level as of the LSCA~\cite{Min10}.
Considering the aforementioned kinematical condition on particles 0 and 1,
one may conclude that for proton both the LSCA and the AMA are valid
in the description of the $^{120}$Sn($p$,$p\alpha$)$^{116}$Cd reaction.
On the other hand, such a validation for particle 2, the knocked
out $\alpha$ particle, has not been done before.

\begin{figure}[htbp]
\begin{center}
 \includegraphics[width=0.45\textwidth]{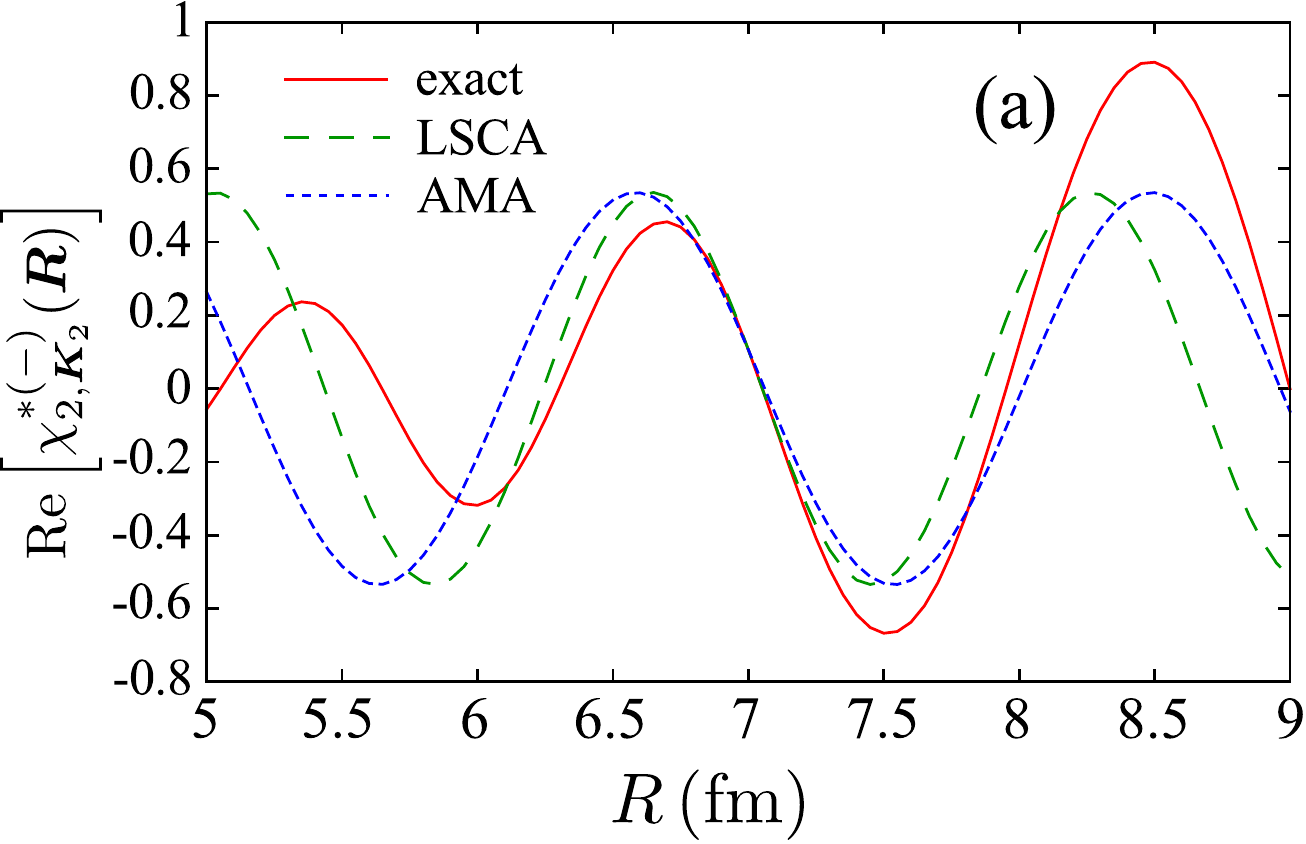}
 \includegraphics[width=0.45\textwidth]{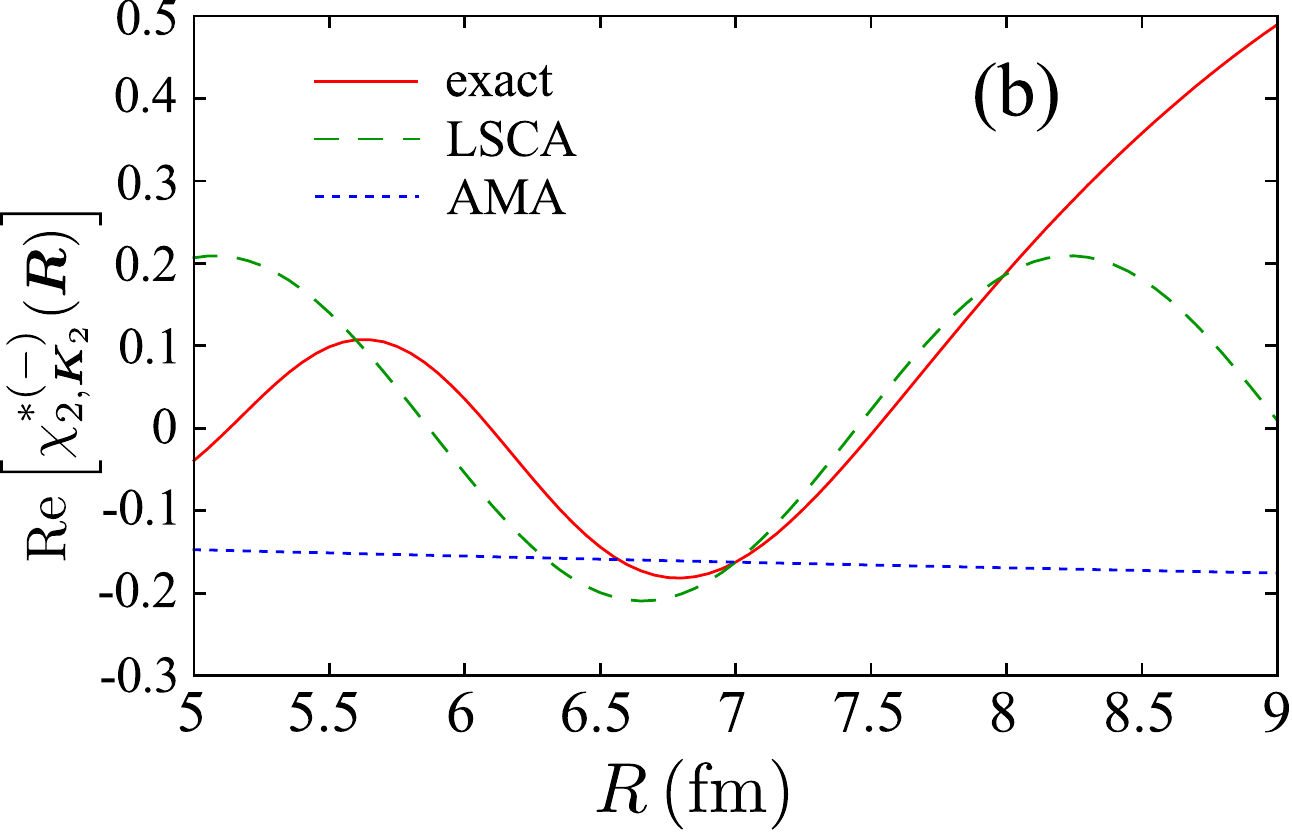}
 \caption{
  (Color online) The test of the LSCA
  and the AMA.
  The real part of $\chi_{2,{\bm K}_2}^{*(-)}$ with no approximation (solid line),
  with the LSCA (dashed line), and with the AMA (dotted line) are plotted.
  In Figs.~\ref{figlsca}(a) and \ref{figlsca}(b), the propagation from
  $(7~{\rm fm}$, $61^\circ$, $180^\circ)$
  and $(7~{\rm fm}$, $29^\circ$, $0^\circ)$ are investigated, respectively;
  $(\theta_2,\phi_2)=(61^\circ,180^\circ)$ is chosen for the kinematics
  of the $\alpha$ particle.\\
  }
 \label{figlsca}
\end{center}
\end{figure}

In Fig.~\ref{figlsca} we show
the validity of the LSCA and the AMA
for $\chi_{2,{\bm K}_{2}}^{(-)}$
with $(\theta_2,\phi_2)=(61^\circ,180^\circ)$,
which corresponds to the quasi-free condition, i.e., the residual nucleus
$^{116}$Cd is at rest in the L frame.
Figures~\ref{figlsca}(a) and \ref{figlsca}(b) correspond to the
propagation from ${\bm R}_a\equiv(7~{\rm fm}$, $61^\circ$, $180^\circ)$
and ${\bm R}_b\equiv(7~{\rm fm}$, $29^\circ$, $0^\circ)$, respectively,
in the spherical coordinate representation.
In each panel the solid, dashed, and dotted lines
show, respectively, the real part of the exact wave function, that with the
LSCA, and that with the AMA.
Since ${\bm R}_a$ (${\bm R}_b$) corresponds
to the foreside (left side) of $^{116}$Cd with respect to the outgoing $\alpha$,
the distortion effect on $\alpha$ at ${\bm R}_a$ (${\bm R}_b$) is
weak (strong).

With weak distortion, as shown in Fig.~\ref{figlsca}(a),
both approximations work well within about 0.5~fm of the propagation.
It should be noted that, with considering the range of the $p$-$\alpha$
interaction of about 2~fm and the constant $1/(A_\alpha+1)=1/5$ in front
of ${\bm s}$ in Eq.~(\ref{r2}), the LSCA and the AMA are required
to be valid for the propagation of about 0.4~fm.
The two approximations are thus validated for the propagation from
${\bm R}_a$.
In case of the strong distortion, as shown in Fig.~\ref{figlsca}(b)
and suggested in Ref.~\cite{Sam86}, the AMA cannot describe the behavior
of the exact scattering wave function;
since the radial direction from ${\bm R}_b$ is almost orthogonal to the
direction of the asymptotic momentum ${\bm K}_2$,
the dotted line is almost constant,
whereas the solid line shows clear variation.
On the other hand, the LSCA reproduces well the exact solution
at almost the same level as in the case of weak distortion.
Thus, one sees that the kinematics of $\alpha$ at ${\bm R}_b$
is significantly changed from that in the asymptotic region by the
distorting potential of $^{116}$Cd; this kinematical change is well
traced by using the LSCA, i.e., the local momentum of the $\alpha$
particle.

Therefore one can conclude that the LSCA works for the $\alpha$
scattering wave function that is strongly distorted, whereas
the AMA not. This may cast doubt on the
use of the factorization approximation for the $(p,p\alpha)$ reaction
investigated in the present study.
In the following subsections we discuss this in view of the TDX.

\subsection{TDX for the $^{120}$Sn($p$,$p\alpha$)$^{116}$Cd reaction at 392~MeV}

\begin{figure}[htbp]
\begin{center}
 \includegraphics[width=0.45\textwidth]{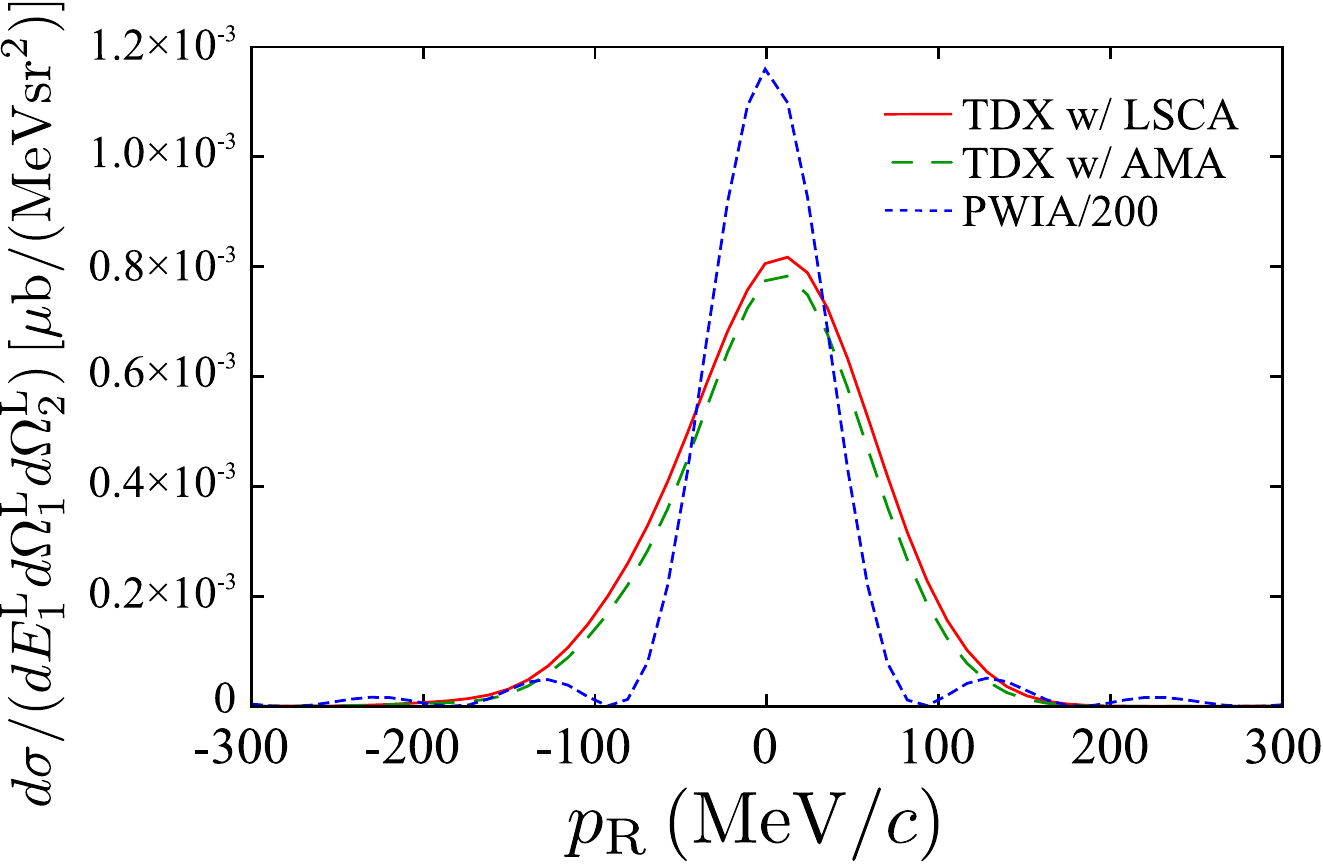}
 \caption{
 (Color online)
 TDX as a function of the recoil momentum.
 The solid (dashed) line corresponds
 to the calculation without (with)
 the factorization approximation.
 The TDX calculated with the PWIA divided by 200 is also shown
 by the dotted line.
 }
 \label{figmd}
\end{center}
\end{figure}

The calculated TDX is shown in Fig.~\ref{figmd} as a function of
the recoil momentum $p_{\rm R}$ defined by
\begin{align}
p_{\rm R}=\hbar K_{\rm B}^{\rm L}\frac{K_{{\rm B}z}^{\rm L}}{|K_{{\rm B}z}^{\rm L}|}.
\end{align}
The solid and dashed lines represent the results without and with the
factorization approximation, respectively.
One sees from the good agreement between the solid and dashed lines that
the factorization approximation, or equivalently, the AMA,
affects very little the TDX, although the AMA for $\alpha$
is shown to be invalid around ${\bm R}_{b}$.
This is due to the strong absorption of $\alpha$ in that region
as shown in Sec.~\ref{sec_ana_integ}.

The dotted line in Fig.~\ref{figmd} represents
the result of the
plane wave impulse approximation (PWIA) calculation divided by
200. The renormalization factor $1/200$ shows
the strong absorption
mainly caused by the $\alpha$-$^{116}$Cd distorting potential.
In the PWIA, the TDX is essentially proportional to the absolute
square of the Fourier transform of the $\alpha$ distribution $\varphi_{\alpha}^{nljm}$
inside $^{120}$Sn. Since we take a 4$S$ state, the dashed line in
Fig.~\ref{figrinteg} shown below, the TDX calculated with the PWIA shows an oscillation
pattern accordingly. The shape of the TDX calculated with the DWIA
is quite different from that with the PWIA. The widening of the
width of the TDX caused by distortion suggests that,
because of the uncertainty principle, only a limited region
of $\varphi_{\alpha}^{nljm}$ is probed by the $(p,p\alpha)$
reaction, as shown in Sec.~\ref{sec_ana_integ}.
It should be noted that
the slight shift of the peak of the TDX with the DWIA from $p_{\rm R}=0$
is understood by the shift of the momentum of particles 2 due to
the real part of the distorting potential~\cite{Oga15}.

\subsection{Probed region of $\alpha$ in $^{120}$Sn by the ($p$,$p\alpha$) reaction}
\label{sec_ana_integ}

\begin{figure}[htbp]
\begin{center}
 \includegraphics[width=0.45\textwidth]{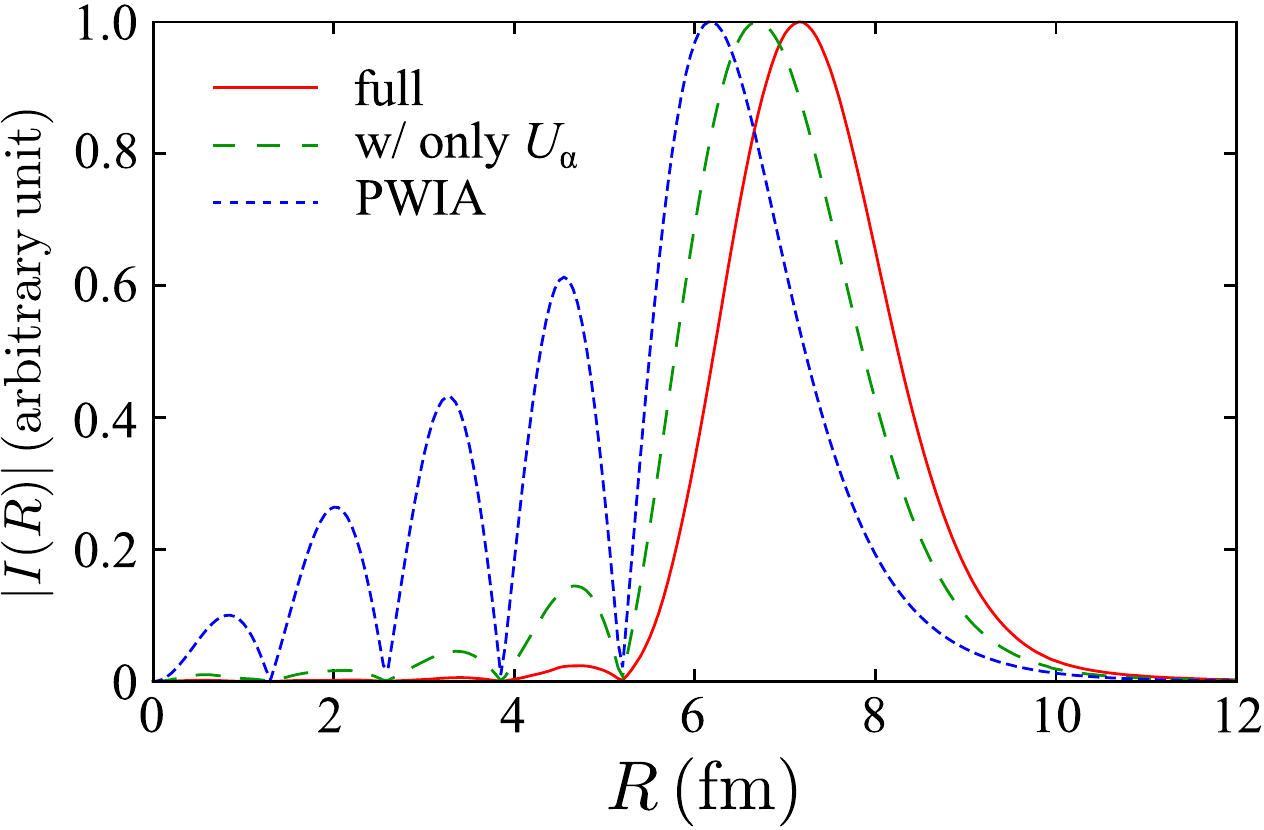}
 \caption{
 (Color online) $|I(R)|$ at
 $p_{\rm R}=0$ (solid line),
 the same but calculated with only the $\alpha$-$^{116}$Cd distorting potential $U_\alpha$
 (dashed line), and the result with PWIA (dotted line). The results are normalized
 to unity at the peak position.
 }
 \label{figrinteg}
\end{center}
\end{figure}

In Fig~\ref{figrinteg}, we show by the solid line
the absolute value of the integrand
on the r.h.s. of Eq.~(\ref{localtmt}) after
integrated over the solid angle $\Omega$ of ${\bm R}$:
\begin{align}
I(R)\equiv&\int
d\Omega\,R^2
\sqrt{
\frac{d\sigma _{p\alpha}}{d\Omega _{p\alpha}}
(\theta _{p\alpha}({\bm R}),E_{p\alpha}({\bm R}))
}
\nonumber \\
&\times \,F_{{\bm K}_0 {\bm K}_1 {\bm K}_2}({\bm R})\,\varphi_{\alpha}^{njlm}({\bm R});
\end{align}
the plotted result corresponds to $p_{\rm R}=0$,
i.e., the quasi-free condition.
The dashed line shows $|I(R)|$ calculated with including only $U_\alpha$, the distorting potential
of the $\alpha$-$^{116}$Cd system in the final state,
and the dotted line shows that with PWIA. Each line is normalized to unity at the peak position.
One sees that
the magnitude of $I(R)$ is strongly suppressed in
the interior region,
$R \la 6$~fm,
mainly because of the absorption due to
the $\alpha$-$^{116}$Cd distorting potential.
The slight shift of the peak position is due to the 
suppression
in the interior region.
It should be noted, however, that the product of the oscillating
three distorted waves and a bound-state wave function can make nontrivial
cancellation. This property also may contribute to the aforementioned
suppression.

\begin{figure}[htbp]
\begin{center}
 \includegraphics[width=0.45\textwidth]{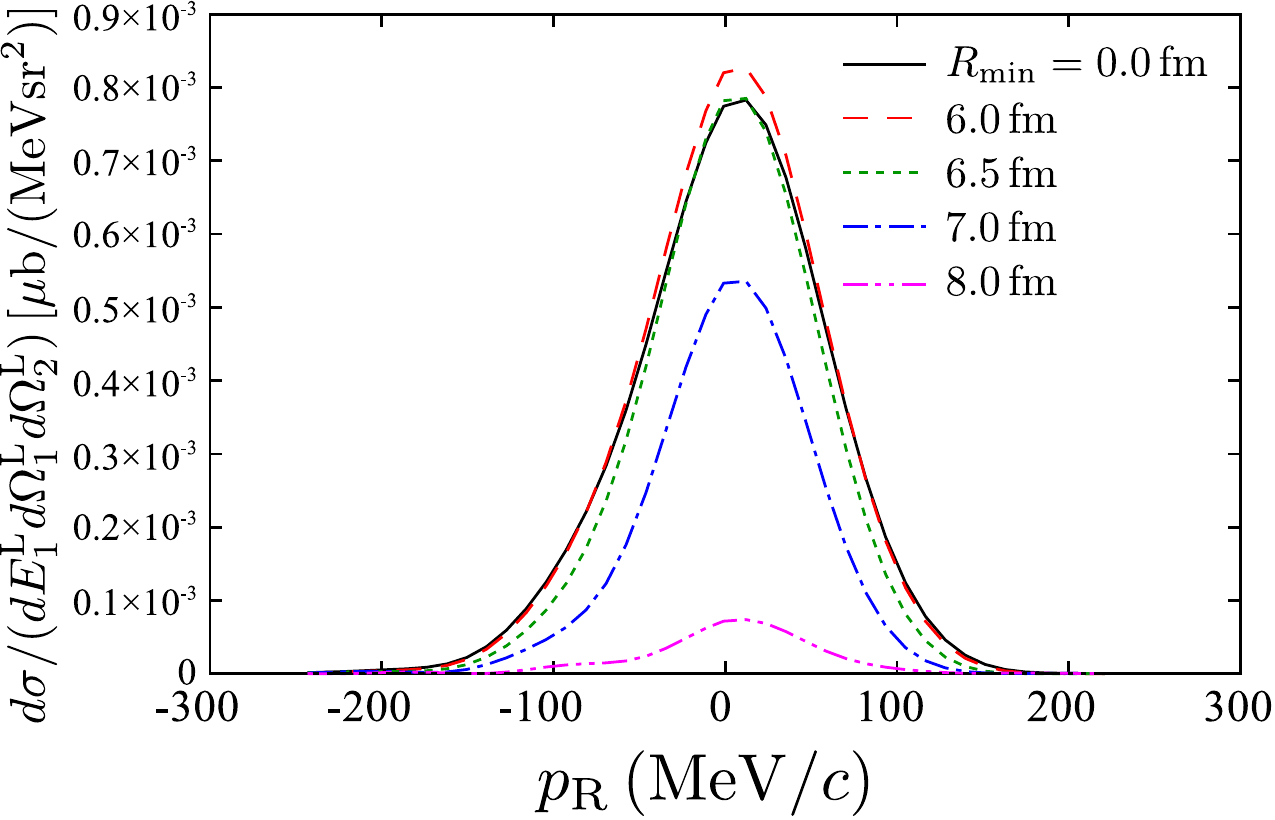}
 \caption{
 (Color online)
 Same as Fig.~\ref{figmd} but with changing $R_{\rm min}$.
 The solid, dashed, dotted, dot dashed and two-dot dashed lines correspond to
 $R_{\rm min} =$~0, 6, 6.5, 7, and 8 fm, respectively.
 }
 \label{figrmin}
\end{center}
\end{figure}

Furthermore, in Fig.~\ref{figrmin}
the TDXs calculated with changing the minimum value $R_{\rm min}$ of the integration over $R$ are shown; we take $R_{\rm min} =$~0, 6, 6.5, 7, and 8 fm.
It is found that the calculated TDX does not change for
$R_{\rm min}=$~0--5.5 fm, and decrease drastically for $R_{\rm min}=$~6--8 fm.
The slight increase of TDX with $R_{\rm min}=6$ fm is due to the
interference of the integrand.
This result shows that the ($p$,$p\alpha$) reaction on heavy nuclei probes
the $\alpha$-cluster wave function on the nuclear surface with
high selectivity, as required for the reaction
to be a good probe for $\alpha$-clustering.
With this peripherality of the ($p$,$p\alpha$)
reaction, one can understand naturally the mechanism that
makes the width of the TDX wider when the distortion is taken into
account.

\begin{figure*}[htbp]
\begin{minipage}{0.33\textwidth}
\begin{center}
 \includegraphics[width=1\textwidth]{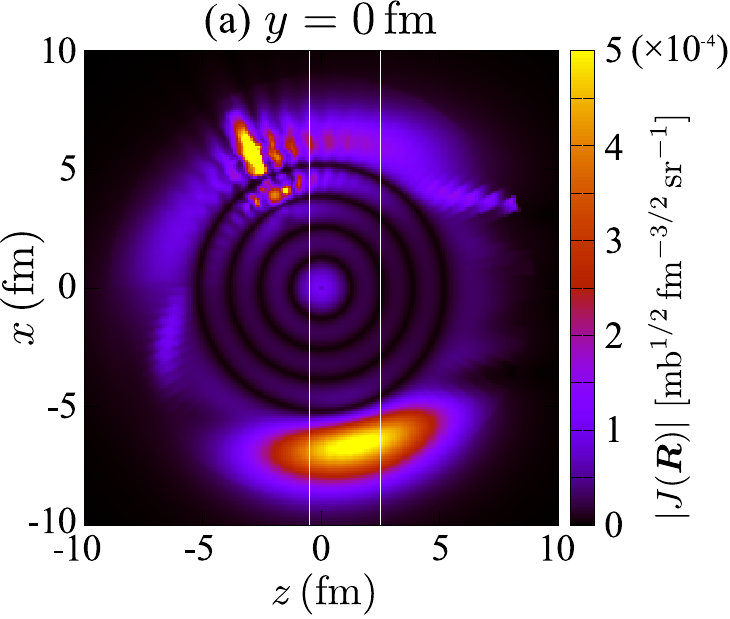}
\end{center}
\end{minipage}
\begin{minipage}{0.33\textwidth}
\begin{center}
 \includegraphics[width=1\textwidth]{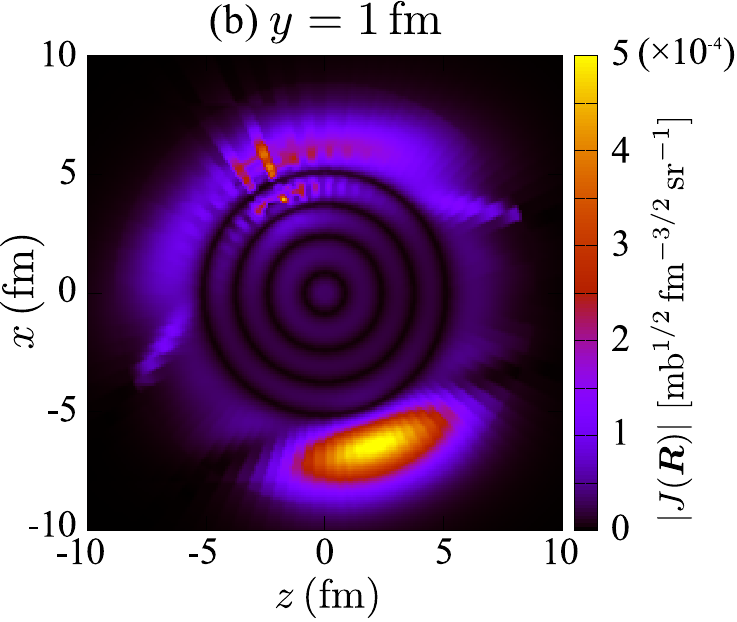}
\end{center}
\end{minipage}
\begin{minipage}{0.33\textwidth}
\begin{center}
 \includegraphics[width=1\textwidth]{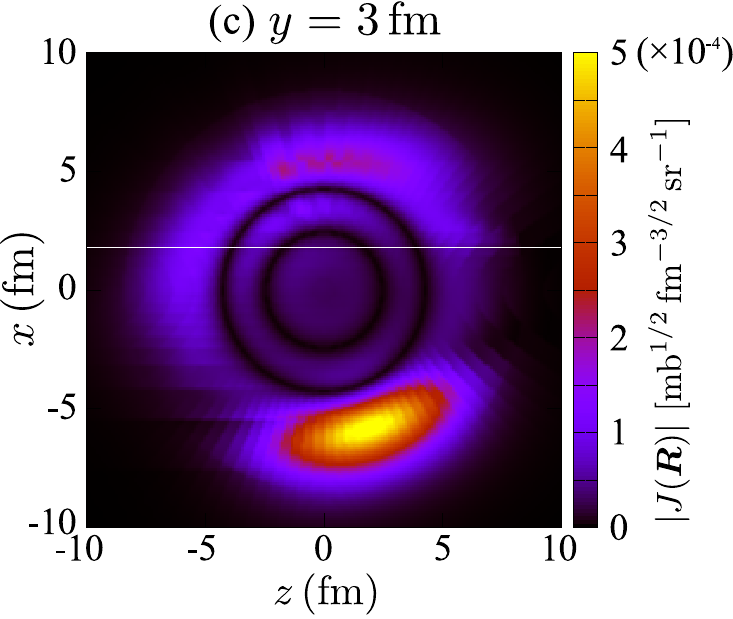}
\end{center}
\end{minipage}
\begin{minipage}{0.33\textwidth}
\begin{center}
 \includegraphics[width=1\textwidth]{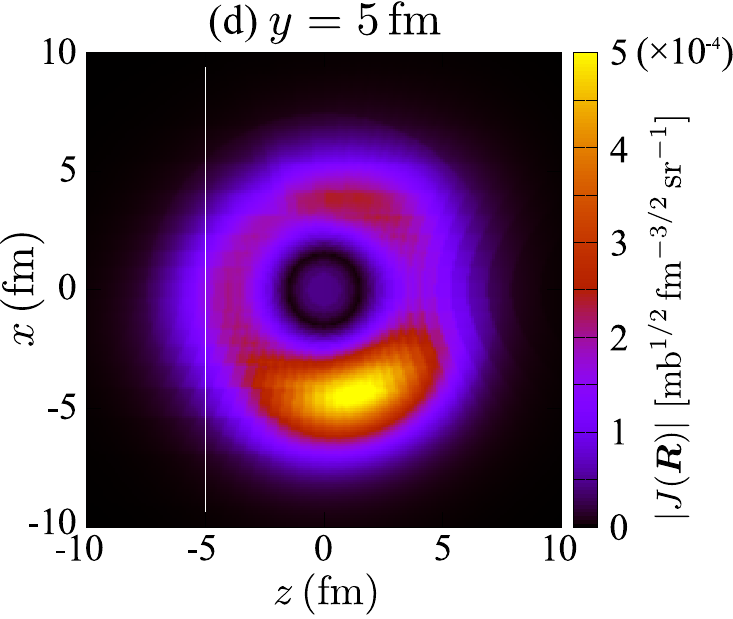}
\end{center}
\end{minipage}
\begin{minipage}{0.33\textwidth}
\begin{center}
 \includegraphics[width=1\textwidth]{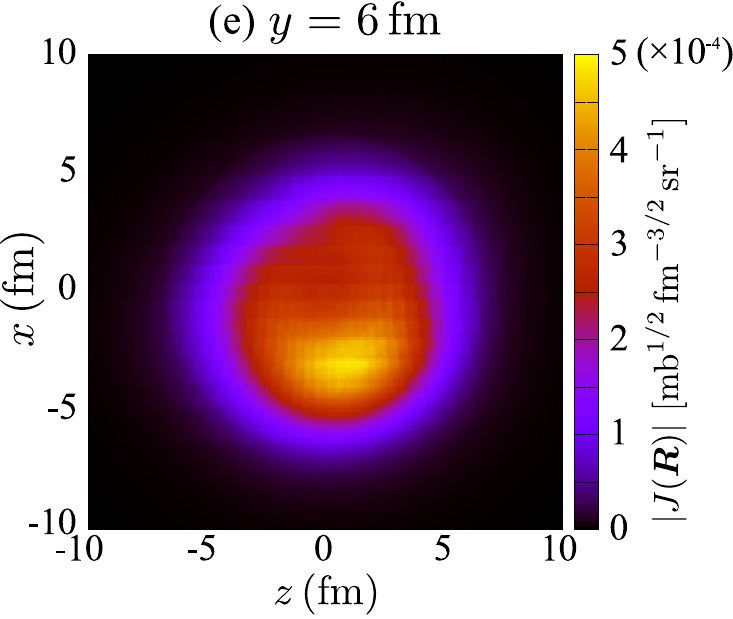}
\end{center}
\end{minipage}
\begin{minipage}{0.33\textwidth}
\begin{center}
 \includegraphics[width=1\textwidth]{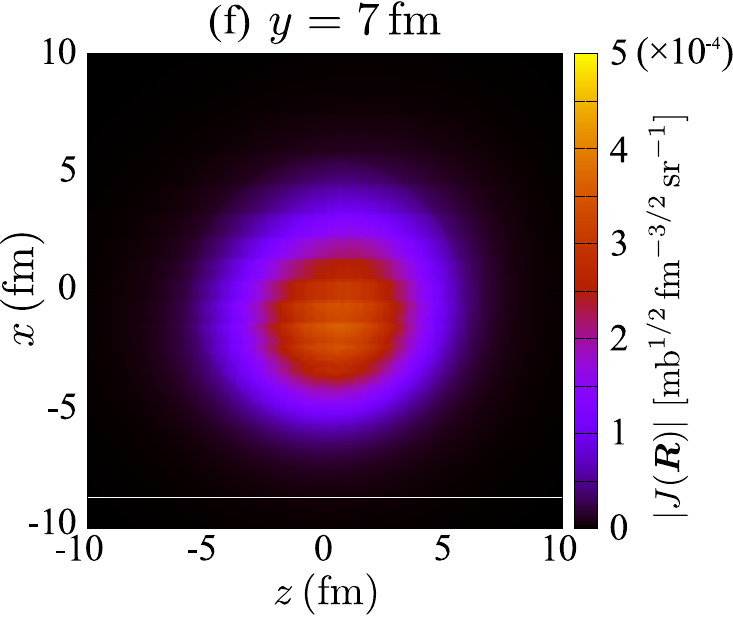}
\end{center}
\end{minipage}
 \caption{
 (Color online)
  $|J({\bm R})|$
 on the $z$-$x$ plane for $y=$0, 1, 3, 5, 6, and 7 fm.
 The kinematical condition is the same as in Fig.~\ref{figrinteg}.
 }
 \label{figrinteg2d}
\end{figure*}

For more detailed analysis, the absolute value of the integrand
on the r.h.s. of Eq.~(\ref{localtmt})
\begin{align}
J({\bm R})\equiv
&
\sqrt{
\frac{d\sigma _{p\alpha}}{d\Omega _{p\alpha}}
(\theta _{p\alpha}({\bm R}),E_{p\alpha}({\bm R}))
}
\nonumber \\
&\times F_{{\bm K}_0 {\bm K}_1 {\bm K}_2}({\bm R})
\varphi_{\alpha}^{nljm}({\bm R})
\label{eqinteg2d}
\end{align}
on the $z$-$x$ plane for $y=$0, 1, 3, 5, 6, and 7 fm are shown in
Fig.~\ref{figrinteg2d}(a)--Fig.~\ref{figrinteg2d}(f).
For $y=$ 0, 1, and 3 fm, it is clearly seen that the amplitude is located in the
foreside region with $R=6$--$9$~fm,
where $\chi_{2,{\bm K}_2}^{(-)}({\bm R})$ is not
absorbed and $\varphi_{\alpha}({\bm R})$
has a finite amplitude.
For $y\ge 5$~fm, the localization of the amplitude becomes rather vague,
because the absorption property of $\chi_{2,{\bm K}_2}^{(-)}({\bm R})$
does not strongly depend on $z$ and $x$ for such values of $y$.
Nevertheless, one may see that the main part of $|J({\bm R})|$ exist
in the foreside region.
Figures~\ref{figrinteg2d}(a)--\ref{figrinteg2d}(f) therefore show that the $(p,p\alpha)$ reaction
has selectivity not only in the radius but also the direction of the
target nucleus.

It is found that the peak at the rear side on $y=0$ plane, around ${\bm R}=(6$--$8~{\rm fm}$, $120^\circ$, $0^\circ)$ in Fig.~\ref{figrinteg2d}(a)
comes from the focus of $\chi_{2,{\bm K}_2}^{*(-)}$ due to the
attraction of the distorting potential
and the increase in $d\sigma_{p\alpha}/d\Omega_{p\alpha}$ caused by that.
It should be noted that this
rear-side peak exists only at around $y=0$ as shown in Fig.~\ref{figrinteg2d},
and makes no major contribution to the TDX.
In fact, it is found that about 90\% of the TDX comes from the
$x < 0$ region. This means that the possible
interference between the amplitudes in the foreside and rear-side regions
is very small, which realizes an intuitive picture that the ($p,p\alpha$)
reaction of our interest takes place in a limited region of space.
These features support that the AMA is valid
for the calculation of the TDX.

\subsection{Discussion of $\alpha$-cluster wave function}
Since a very naive model for $\varphi_{\alpha}$ is adopted in the present study,
it is important to see the $\varphi_{\alpha}$ dependence of the findings discussed above.
It is obvious that the validity of the LSCA itself has nothing to do with $\varphi_{\alpha}$. Thus, we discuss the $\varphi_{\alpha}$ dependence
of the TDX as well as effect of the AMA on that.
\begin{figure}[htbp]
\begin{center}
 \includegraphics[width=0.45\textwidth]{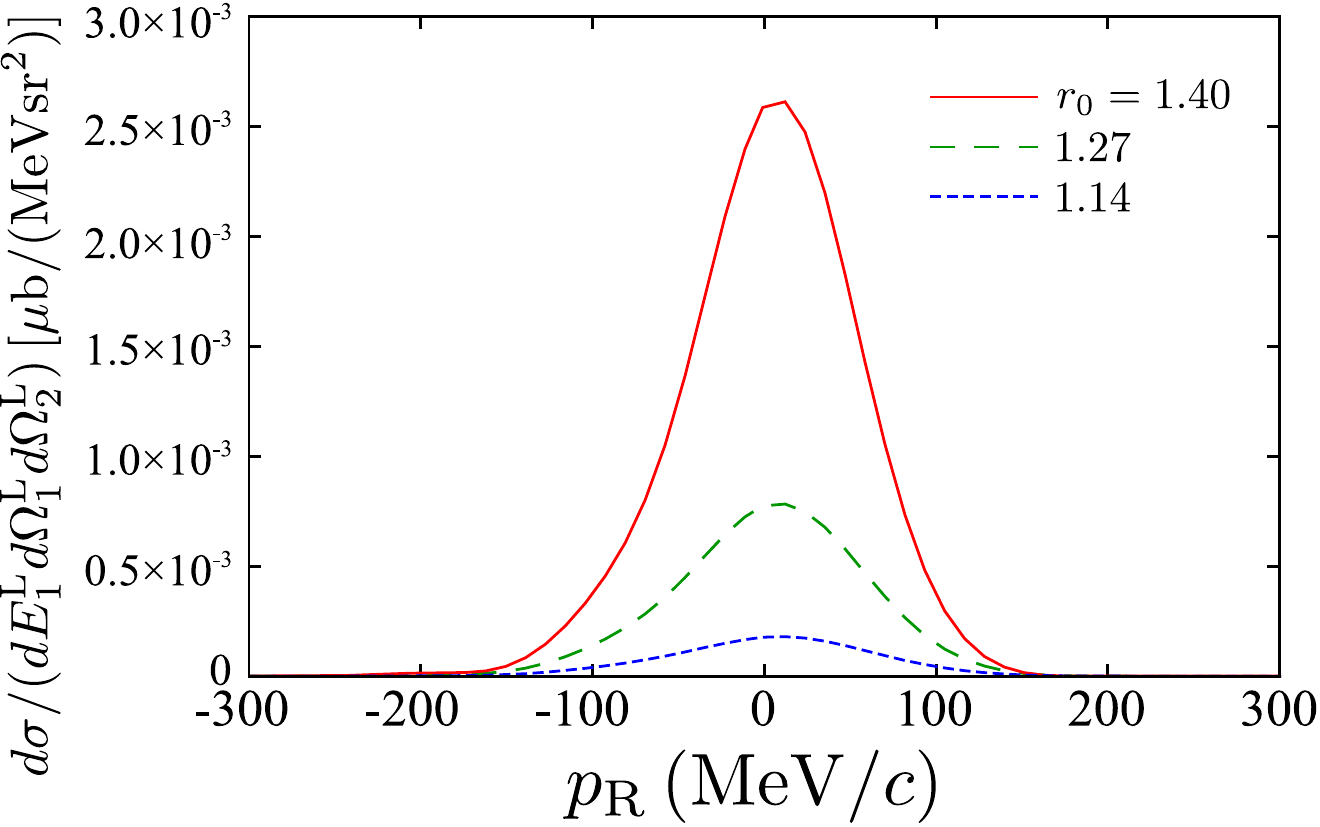}
 \caption{
 (Color online)
 The TDXs with different $r_0$.
 The solid (dotted) line is the TDX with $r_0=1.40$ ($1.14$) fm.
 The dashed line is the same as Fig.~\ref{figmd} for comparison.
  }
 \label{figrange}
\end{center}
\end{figure}
In Fig.~\ref{figrange} the solid (dotted) line shows the TDX calculated with $\varphi_{\alpha}$ with increasing (decreasing) the range parameter
$r_0$ by 10\%,
$r_0=1.40$ ($1.14$)~fm; these results are obtained by using the LSCA.
The dashed line is the same as the solid line in Fig.~\ref{figmd}.
One can see that the 10\% difference of $r_0$ changes the magnitude of the
TDX significantly, i.e., about a factor of three difference.
This is also understood by the absorption in the interior region.
Since only the surface region contributes to the TDX,
small extension of $\varphi_{\alpha}$ to the exterior region
changes the magnitude of TDX drastically.
It is found that
the TDX at $p_{\rm R}=0$ calculated with the AMA differs from that with
the LSCA by only 6\% at most.
Furthermore, the qualitative features shown in Figs.~\ref{figrmin} and
\ref{figrinteg2d} turned out to be independent of $r_0$.

\section{Summary}
\label{secsum}
We have examined the $^{120}$Sn($p$,$p\alpha$)$^{116}$Cd reaction at
392~MeV in the DWIA framework.
To show the validity of the DWIA model, we have demonstrated that
it reproduces the observed energy sharing cross section data of
$^{66}$Zn($p,p\alpha$)$^{62}$Ni at 101.5 MeV.
It was clarified that the so-called factorization approximation
adopted in many preceding studies is equivalent to the AMA to the
distorted waves, which is a further simplification of the LSCA.
Although the AMA does not work for the propagation of $\alpha$
in the region where the nuclear deflection is significant,
it does not affect the TDX because of the strong
absorption in that region.
In other words, the integrand of the transition matrix has a
contribution only in the region where the AMA works well.
As a result, the factorization approximation
was verified for the calculation of the TDX of the $(p,p\alpha)$
reaction.
It should be kept in mind, however, that the inaccuracy of the AMA
may affect the TDX if a scattering particle feels a potential
having a strong real part and a weak imaginary part;
this can be realized, for instance, for nucleon scattering at
lower energies.
The strong absorption due to the $\alpha$-$^{116}$Cd distorting
potential makes the $(p,p\alpha)$ reaction very peripheral, which
allows one to clearly probe the $\alpha$-clustering
of nuclei.
Furthermore, the $(p,p\alpha)$ reaction has
high selectivity also in the direction of the target nucleus; only the
foreside region with respect to the emitting $\alpha$ with
the radius of 6--9~fm is probed.
It is also shown that the factorization approximation and the peripherality
of the reaction are valid for
different choices of $\varphi_{\alpha}$, but the magnitude of TDXs are
strongly dependent on them.
This result suggests that it is essential to employ a reliable
alpha-cluster wave function for the qualitative discussion.

Validation of
the on-shell approximation to the
$p$-$\alpha$ transition amplitude will be
important for
more reliable description
of the knockout processes.

\section*{ACKNOWLEDGMENTS}
The authors thank S.~Kawase and T.~Uesaka for
fruitful discussions.
The computation was carried out with the computer facilities
at the Research Center for Nuclear Physics, Osaka University.
This work was supported in part by Grants-in-Aid of the Japan Society
for the Promotion of Science (Grants No. JP15J01392 and No. JP25400255)
and by the ImPACT Program of the Council for Science, Technology and
Innovation (Cabinet Office, Government of Japan).

%%--------------------------------------------------------------------%%
%%                           References                               %%
%%--------------------------------------------------------------------%%

\end{document}